\documentclass[aps,prl,twocolumn,groupedaddress]{revtex4}
\usepackage[dvips]{graphics}

\begin{document}


\title{Turbulent Friction in Rough Pipes  and the Energy \\
 Spectrum of the Phenomenological Theory}


\author{G.\ Gioia}
\author{Pinaki Chakraborty}
\affiliation{Department of Theoretical and Applied Mechanics\\ University of Illinois at Urbana-Champaign, Urbana, IL 61801}


\date{\today}

\begin{abstract}
The classical experiments on
turbulent friction in rough pipes
were performed by J.\ Nikuradse in the 1930's.
  Seventy years later, they
 continue to defy theory.
Here we model 
 Nikuradse's experiments using
the phenomenological
theory of Kolmog\'orov,
a theory that is widely thought to be
applicable only to highly idealized flows. 
Our results include both the
empirical scalings of Blasius and 
Strickler, and are otherwise in minute 
qualitative agreement with the
experiments; they suggest that the phenomenological theory 
may be relevant to other flows of
practical interest; and they unveil the
existence of close ties between two milestones
of experimental and theoretical turbulence.

\end{abstract}


\maketitle


 Turbulence is the unrest that 
 spontaneously takes over a streamline
  flow adjacent to a wall or obstacle
 when the flow is 
 made sufficiently fast. Although
 most of the flows that surround us 
  in everyday life and in nature
  are turbulent flows over rough walls, 
  these flows have remained amongst 
 the least understood phenomena of 
 classical physics \cite{jimenez,antonia}.
  Thus, one of the weightier
  experimental studies of turbulent flows 
  on rough walls, and the most useful 
  in common applications, is yet to be
 explained theoretically
 70 years after its publication.
 In that study \cite{niku}, Nikuradse elucidated how
the friction coefficient  
between the wall of a pipe and 
 the turbulent flow inside depends 
 on the Reynolds number of the flow
 and the roughness of the wall. 
 The friction coefficient, $f$, is a measure
 of the shear stress (or shear force per unit area)
 that the turbulent flow exerts on the
 wall of a pipe; it is customarily
 expressed in dimensionless form
 as $f=\tau/\rho V^2$, where $\rho$ is
 the density of the liquid that flows in the pipe
 and $V$ the mean velocity of the flow. 
 The Reynolds number is defined as
 ${\rm Re}= V R/\nu$, where  $R$ is the radius of the pipe
 and $\nu$ the kinematic viscosity of the liquid. 
 Last, the roughness
 is defined as the ratio $r/R$ between
 the size $r$ of the roughness
 elements (sand grains in the case of Nikuradse's experiments)
  that line the wall of
 the pipe and the radius of the pipe.

\begin{figure} 
\centering
\resizebox{3.4in}{!}{\includegraphics{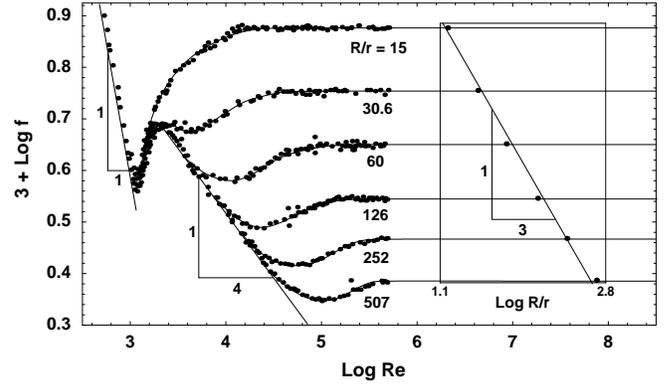}}
\caption{\label{fig1} Nikuradse's data. Up to a ${\rm Re}$
 of about $3,000$
 the flow is streamline (free from turbulence)
  and $f\sim 1/{\rm Re}$. Note that 
 for very rough pipes (small $R/r$) 
 the curves do not form a belly at intermediate values of Re.
 Inset: verification of Strickler's empirical scaling
 for $f$ at high Re, $f \sim (r/R)^{1/3}$.   }
\end{figure}

 Nikuradse presented his data in the form
 of six curves, the log-log plots of $f$ versus Re
  for six values of the roughness \cite{niku}.
 These curves are shown in Fig.~\ref{fig1}.
 At the onset of turbulence \cite{hof}, 
 at a Re of about 3,000,
 all six curves rise united in a single bundle. 
 At a Re of about 3,500, the bundle
 bends downward to form a marked 
 {\it hump\/} and then it plunges in accord 
 with Blasius's empirical scaling \cite{prandtl},
 $f \sim {\rm Re}^{-1/4}$, as 
 one by one in order of decreasing roughness
 the curves start to careen away from the 
 bundle.  After leaving the bundle, 
 which continues to plunge,
 each curve sets out to trace a {\it belly\/} \cite{belly}
 as it steers farther from the bundle with increasing Re,
 then flexes towards 
 a terminal, constant value of $f$ that is in keeping
 with Strickler's empirical scaling \cite{strickler},
 $f \sim (r/R)^{1/3}$. 
 For seventy years now,
 our understanding of these curves
 has been aided by little beyond 
 a pictorial narrative
 of roughness elements being progressively
 exposed to the turbulent flow as Re increases \cite{chow}.

In our theoretical work, we adopt the
 phenomenological imagery of 
 ``turbulent eddies'' \cite{rich,kolmo,frisch}
 and use the 
 spectrum of turbulent energy \cite{pope} at a lengthscale $\sigma$, 
 $E(\sigma)$, to determine the velocity of 
 the eddies of size $s$, $u_s$, in the form
$u_s^2=\int_0^s E(\sigma) \sigma^{-2} d\sigma$, where 
$E(\sigma)=A\,\varepsilon^{2/3}\sigma^{5/3}c_d(\eta/\sigma)c_e(\sigma/R)$.
  Here $A$ is a dimensionless constant,
   $\varepsilon$ is the turbulent power per unit mass,
 $\eta=\nu^{3/4}\varepsilon^{-1/4}$ is the 
 viscous lengthscale, $R$ is the largest lengthscale in the flow, 
$A\,\varepsilon^{2/3}\sigma^{5/3}$ is
 the Kolmog\'orov spectrum (which is valid in the inertial range, 
 $\eta\ll\sigma\ll R$), and $c_d$ and $c_e$ are dimensionless 
 corrections for 
 the dissipative range and the energetic range, 
 respectively. For $c_d$ we adopt an exponential
 form, $c_d(\eta/\sigma)=\exp(-\beta\eta/\sigma)$ (which gives 
 $c_d\approx 1$ except in the dissipative range,
 where $\sigma\approx\eta$), and 
  for $c_e$ the form proposed by von K\'arm\'an,
 $c_e(\sigma/R)=(1+\gamma(\sigma/R)^2)^{-17/6}$ (which gives 
 $c_e\approx 1$ except in the energetic range,
 where $\sigma\approx R$), where $\beta$ and $\gamma$ are
  dimensionless constants \cite{pope}. 
 To obtain expressions for $u_s$ and $\eta$ in terms of 
 Re, $r/R$, and $V$, we invoke the usual scalings \cite{tl}, 
 $\varepsilon=\kappa_\varepsilon\,u_R^3/R$ 
 (Taylor's scaling \cite{taylor}, 
 where $u_R$ is the characteristic velocity of the largest eddies
 and $\kappa_\varepsilon$ a dimensionless constant) and 
 $u_R=\kappa_u V$ (where $\kappa_u$ is a dimensionless constant).
 Then, we can write $\eta=b R\,{\rm Re}^{-3/4}$, where 
 $b\equiv (\kappa_\varepsilon \kappa_u^3)^{-1/4}$, and (after
 changing the integration variable to $x\equiv\sigma/R$)
 $u_s^2=A\, \kappa_\varepsilon^{2/3}  
    u_R^2\int_0^{s/R}x^{-1/3}c_d(b\,{\rm Re}^{-3/4}/x)c_e(x)dx$.
 For $s\ll R$ we can set $c_e=1$, compute the integral, and
 let ${\rm Re}\to\infty$ to obtain
 $u_s^2 =(3/2) A\, \kappa_\varepsilon^{2/3} u_R^2\,(s/R)^{2/3}$,
  or $u_s^2\sim u_R^2 (s/R)^{2/3}$, a well-known result of 
 the phenomenological theory. Further, for consistency with
 Taylor's scaling we must have
 $A\, \kappa_\varepsilon^{2/3}=2/3$ (so that $u_s=u_R$ for $s=R$)
 and therefore 
 $u_s^2= \kappa_u^2\,V^2 (2/3) \int_0^{s/R}x^{-1/3}c_d(b\,{\rm Re}^{-3/4}/x)c_e(x)dx$.

\begin{figure}
\centering 
\resizebox{2.3in}{!}{\includegraphics{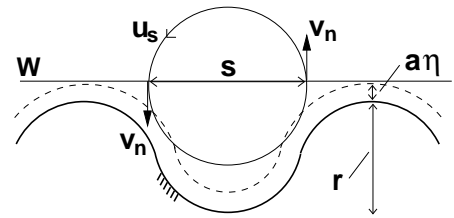}}
\caption{\label{fig2}   Schematic of the
 immediate vicinity of the wall with
  roughness elements of size $r$ covered by a
 viscous layer of uniform thickness $a \eta$.
  The distance between roughness elements
 is about equal to the height of the roughness elements,
 as in Nikuradse's experiments \cite{kd}.
 The horizontal line is the trace of a wetted surface 
 $W$ tangent to the peaks of the viscous layer.
}
\end{figure}

We now seek to derive an expression for
$\tau$, the shear stress on the wall of the pipe. 
 We assume a viscous layer
  of constant thickness $a\eta$,
  where $a$ is a dimensionless constant,
 and call $W$ a wetted surface 
 parallel to the peaks of 
 the viscous layer (Fig.~\ref{fig2}). 
Then, $\tau$ is effected by
 momentum transfer across $W$.
 Above $W$, the velocity of the flow scales
 with $V$, and the fluid carries
 a high horizontal momentum per unit volume
 ($\sim\rho V$). Below $W$, the velocity of the 
 flow is negligible, and the fluid carries
 a negligible horizontal momentum per unit volume.
 Now consider an eddy that straddles the wetted
 surface $W$. This eddy transfers fluid of high
 horizontal momentum downwards across $W$, 
 and fluid of negligible horizontal momentum
 upwards across $W$. 
 The net rate of transfer of momentum across $W$ 
 is set by the velocity normal to $W$,
 which velocity is provided by the eddy. 
 Therefore, if $v_n$ denotes 
 the velocity normal to $W$ provided
  by the {\it dominant\/} eddy that straddles
 $W$, then the shear stress effected by 
 momentum transfer across $W$ scales in the form
 $\tau\sim\rho\,V v_n$.  
 
 In order to 
 identify the dominant eddy that straddles $W$,
 let us denote by $s=r+a\eta$ 
  the size of the largest eddy that
  fits the coves between 
 successive roughness elements.
 Eddies much larger than $s$
  can provide only a negligible velocity
 normal to $W$.
 (This observation is purely a matter of
  geometry.) On the other hand, 
 eddies smaller than  $s$ 
 can provide a sizable velocity normal 
 to $W$. Nevertheless,
  if these eddies are much smaller
 than $s$, their velocities are 
 overshadowed by
 the velocity of the eddy of size $s$. 
 Thus, $v_n$ scales with $u_s$, which is the 
 velocity of the 
 eddy of size $s$, and the dominant eddy
 is the largest eddy that
  fits the coves between 
 successive roughness elements.
  We conclude that 
 $\tau\sim\rho\, V u_s$,
 or  $\tau=\kappa_\tau\rho\, V u_s$
 (where $\kappa_\tau$ is a dimensionless constant of order 1),
and therefore $f=\kappa_\tau u_s/V$ or
\begin{equation}\label{efe}
f=K \!\left(\int_0^{s/R}x^{-1/3} 
c_d(b\,{\rm Re}^{-3/4}/x)c_e(x)dx\right)^{1/2},
\end{equation}
where $K\equiv\kappa_\tau\kappa_u\sqrt{2/3}$,
 $s/R=r/R+ a b\,{\rm Re}^{-3/4}$, and
 $b\equiv (\kappa_\varepsilon \kappa_u^3)^{-1/4}$.
Equation~(\ref{efe})
 gives $f$ as an explicit function of the Reynolds number
 Re and the roughness $r/R$.

\begin{figure}
\centering 
\resizebox{3.4in}{!}{\includegraphics{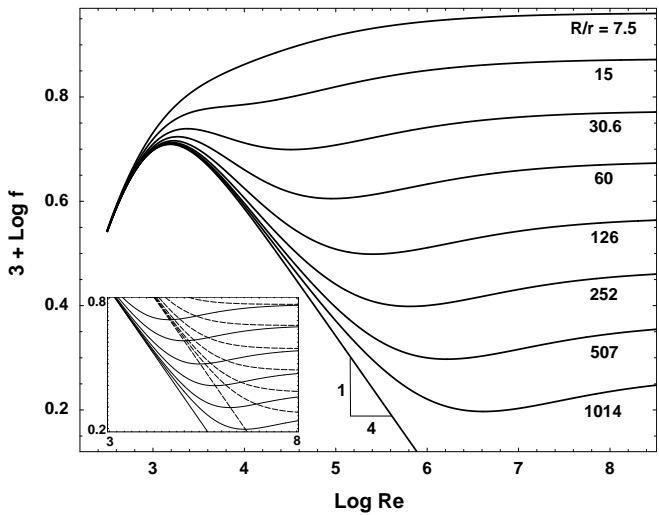}}
\caption{\label{fig3} Plot of (\ref{efe}). 
  Inset: Plot of (\ref{efe1}) 
 (no correction for the energetic range: solid lines)
  and plot of (\ref{efe1}) with $\gamma=0$
 (no correction for the energetic range and 
 the dissipative range: dashed lines).}
\end{figure}

To evaluate
 computationally the integral of (\ref{efe}),
 we set $\beta=2.1$, $\gamma=6.783$ 
 (the values given in \cite{pope}), $a=5$ 
 ($5 \eta$ being a common estimation of the
 thickness of the viscous layer), 
 $\kappa_\varepsilon=5/4$
 (a value that follows from Kolmog\'orov's four-fifth law \cite{ff}),
 $\kappa_u=0.036$ 
 ($0.036\pm 0.005$ being the value measured in pipe flow 
   by Antonia and Pearson \cite{ap}), 
 $b\equiv(\kappa_\varepsilon \kappa_u^3)^{-1/4}=11.4$,
  and treat $\kappa_\tau$ as a free parameter 
 (albeit a parameter constrained by  
 theory to be of order 1). 
 With $\kappa_\tau=0.5$ (and therefore $K=0.015$),
  (\ref{efe}) gives the plots of
 Fig.~\ref{fig3}. (Note that a different value of $\kappa_\tau$ 
 would give the same plots except for a vertical translation.)
  These plots show that (\ref{efe}) is in excellent
 qualitative agreement with Nikuradse's data, right from
 the onset of turbulence, including the hump 
 and, for relatively low roughness, the bellies. 
 These plots remain qualitatively
 the same even if the value of any of the 
 parameters is changed widely. In particular,
   there is always a hump and there are always bellies:
 these are robust features  closely 
 connected with the overall form of the spectrum 
 of turbulent energy. The connections will
  become apparent after the discussion that 
 follows.

To help interpreting our results, 
 we compute $f$  without including the correction 
 for the energetic range---that is, setting $\gamma=0$.
 In this case, the integral of (\ref{efe}) may be
 evaluated analytically, with the result
\begin{equation}\label{efe1}
f=K (r/R + a b\,{\rm Re}^{-3/4})^{1/3} \sqrt{F(y)},
\end{equation}
where 
 $F(y)= y^{2/3}\Gamma_{-2/3}(y)$,
$\Gamma_{-2/3}$ 
 is the gamma function of order $-2/3$, and  
  $y=\beta\eta/s=\beta b\,{\rm Re}^{-3/4}(r/R+a b\,{\rm Re}^{-3/4})^{-1}$.
 With the same values of $\kappa_\tau$, $\kappa_u$,  $a$, $b$,
 and $\beta$ as before,
  (\ref{efe1}) gives the solid-line plots in the inset
 of Fig.~\ref{fig3}. The
 hump is no more. We conclude that the hump relates to
 the energetic range. Further, with the exception
 of the hump at relatively low Re,
 the plots of  (\ref{efe}) coincide with the plots 
 of (\ref{efe1}); thus, we can study (\ref{efe1}) 
 to reach conclusions about (\ref{efe}) at intermediate 
 and high Re. For example, (\ref{efe1}) gives 
 $f\sim(r/R)^{1/3}$ for $r\gg a\eta$  and
 $f\sim{\rm Re}^{-1/4}$ for $r\ll a\eta$. It follows
 that both (\ref{efe1}) and (\ref{efe}) give a gradual 
 transition between the empirical 
 scalings of Blasius and Strickler \cite{gb},
  in accord with Nikuradse's data.

 If we set $\beta=0$ in addition to $\gamma=0$, (\ref{efe1}) 
 simplifies to $f=\kappa_\tau\kappa_u\,(r/R+a b\,{\rm Re}^{-3/4})^{1/3}$.
 With the same values of $\kappa_\tau$, $\kappa_u$, $a$, and $b$ as before,
  this expression gives the dashed-line plots in the inset 
 of Fig.~\ref{fig3}.
 Now the bellies are no more. We conclude that the bellies 
 relate to the dissipative range. The dissipation 
 depresses the values of $f$ at relatively low and
 intermediate Re, leading to the formation of the
  bellies of Nikuradse's data.  

\begin{figure}
\centering 
\resizebox{3.38in}{!}{\includegraphics{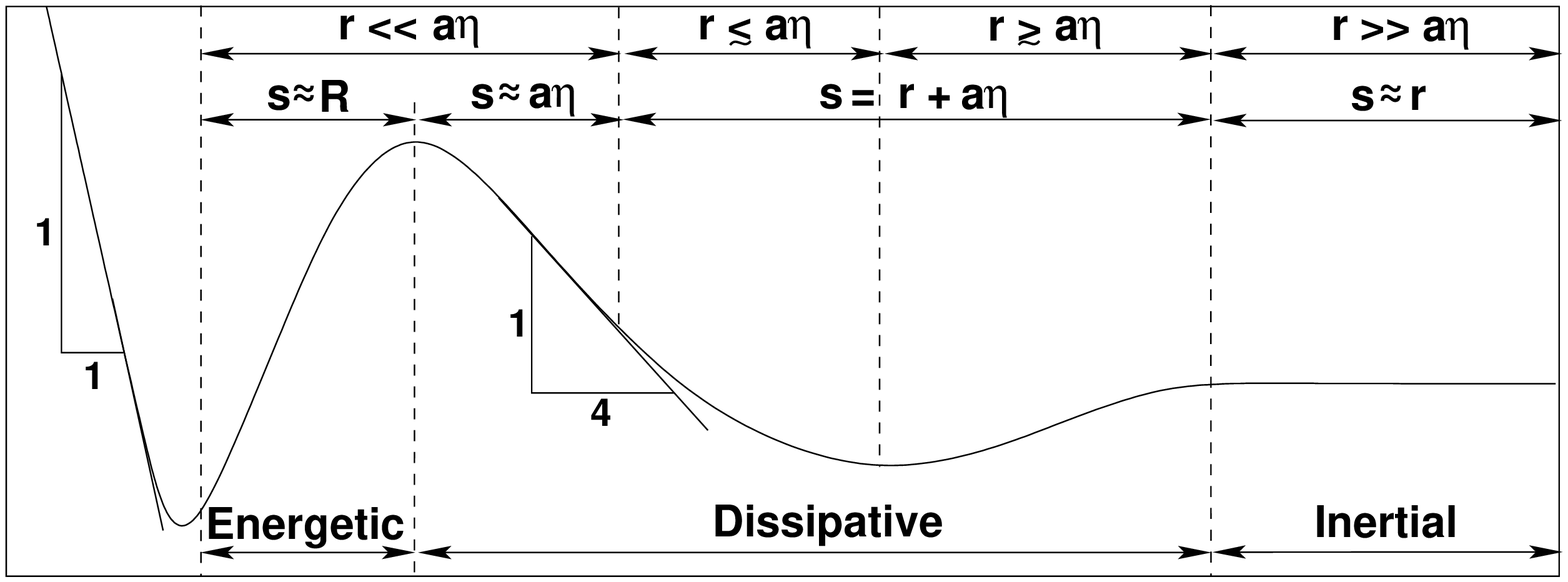}}
\caption{\label{fig4} Schematic of the relations among
 a generic Nikuradse curve, the spectrum of turbulent energy,
 the size of the roughness elements, the thickness of the
 viscous layer, and the size of the dominant eddies.  
}
\end{figure}

 We are ready to explain the unfolding 
 of Nikuradse's data in terms of the varying  
 habits of momentum transfer with increasing Re
  (Fig.~\ref{fig4}).
At relatively low Re, the inertial range 
 is immature, and the momentum transfer
 is dominated by eddies in the
 energetic range, whose velocity scales with 
 $V$, and therefore with Re.
 Consequently, an increase in Re leads to a more 
 vigorous momentum transfer---and to an increase in
 $f$. This effect explains the rising part of the
 hump. At higher Re,
the momentum transfer is dominated
 by eddies of size $s\approx a\eta\gg r$.
 Since $\eta\sim{\rm Re}^{-3/4}$, with increasing
 Re the momentum transfer is effected by ever 
 smaller (and slower) eddies, and $f$ lessens as Re continues 
 to increase. This effect explains the plunging 
 part of the hump---the 
 part governed by Blasius's scaling. At 
 intermediate Re, $s=r+a\eta$ with
 $r\approx a\eta$. Due to the decrease in $\eta$,
 $s$ continues to lessen as Re 
 continues to increase, but at a lower
 rate than before, when it was 
 $s\approx a\eta\gg r$.
 Thus, the curve  
 associated with $r$ 
 deviates from Blasius's scaling
 and starts to trace a belly. 
 As $\eta$ continues to decrease,
  the dominant eddies
 become decidedly 
 larger than the smaller eddies
  in the inertial range, which is
  well established now, and any 
 lingering dissipation 
 at lengthscales larger than $s$
 must cease. This effect explains 
 the rising part of the belly.
 Last, at high Re, 
  $s\approx r\gg a\eta$.
 As Re increases further, 
 $\eta$ lessens and new, smaller
 eddies populate the flow and become jumbled with
 the preexisting eddies. Yet the momentum transfer
 continues to be dominated by eddies of
 size $r$, and $f$ remains invariant. 
 This effect explains Nikuradse's data
  at high Re, where $f$ is
 governed by Strickler's scaling. 

 We have predicated equation (\ref{efe}),
 on the assumption that the turbulent 
  eddies are governed 
 by the phenomenological
  theory of turbulence. The theory 
 was originally derived for 
  isotropic and homogeneous flows,
  but recent research \cite{aniso}
  suggests that it
  applies to much more general flows as well.
  Our results
 indicate that even where the flow is 
 anisotropic and inhomogeneous---as is 
 the case in the vicinity of a wall---the
 theory gives an approximate 
 solution that embodies the essential structure of 
 the complete solution (including 
  the correct scalings of Blasius and Strickler) and 
 is in detailed qualitative agreement with the
 observed phenomenology. Remarkably, 
  the qualitative agreement holds starting at
 the very onset of turbulence, in accord
 with experimental evidence that 
 ``in pipes, turbulence sets in  suddenly
 and fully, without intermediate states
 and without a clear stability boundary'' \cite{hof}.
The  deficiencies in quantitative agreement 
 point to a need for 
 corrections 
 to account for the effect of 
 the roughness elements on the 
 dissipative range
 as well as for the effect of the overall 
 geometry on the energetic range.


 In conclusion, to a good approximation 
 the eddies in a pipe are governed
 by the spectrum of turbulent energy of
 the phenomenological theory. The size of
 the eddies that dominate the momentum 
 transfer close to the wall 
 is set by a combination of the size of 
 the roughness elements and 
 the viscous lengthscale. As a result,
  the dependence of the
 turbulent friction on the roughness and the
 Reynolds number is a direct manifestation of
 the distribution of turbulent energy
 given by the 
 phenomenological theory. This close
  relation between the turbulent friction 
 and the phenomenological theory \cite{ng} may be summarized
 in the following observation: the similarity
 exponents of Blasius and Strickler 
 are but recast forms of the 
 exponent $5/3$ of the Kolmog\'orov spectrum.

\begin{acknowledgments}
We are thankful to
 F.\ A.\ Bombardelli, N.\ Goldenfeld, and W.\ R.\ C.\ Phillips
 for a number of illuminating discussions.
 We are also thankful to Referee C, whose pointed criticism
  resulted in a much stronger paper.
 J.\ W.\ Phillips kindly read our manuscript and made
 suggestions for its improvement. 
\end{acknowledgments}


\begin{thebibliography}{} 
\bibitem{jimenez}{J.\ Jim\'enez, 
 Annu.\ Rev.\ Fluid Mech.\ {\bf 36}, 173 
(2004).} 
\bibitem{antonia}{M.\ R.\ Raupach, R.\ A.\ Antonia,
S.\ Rajagopalan, 
 Appl.\ Mech.\ Rev.\ {\bf 44}, 1 
(1991).} 
\bibitem{niku}{Reprinted in English in 
   J.\ Nikuradse, 
   {\it NACA TM} {\bf 1292} (1950).
} 
\bibitem{hof}{B.\ Hof et al.,
  Science {\bf 305}, 1594 (2004).}
\bibitem{prandtl}{L.\ Prandtl, {\it Essentials of Fluid Dynamics\/}, 
    (Blackie \& Son, London,  3rd ed., 1953), chap. III.11.}
\bibitem{belly}{In Nikuradse's 
 experiments the average distance between roughness
 elements, $\lambda$, was about the same as the height 
 of the roughness elements, $r$. This is the type of 
 single-lengthscale roughness that concerns us here. 
 For this type of roughness, there are always
 bellies in the log-log plots of $f$ vs.\ Re.
 (For similar results on open channels,
 see Varwick's data in, e.g., O.\ Kirshmer,
  Revue g\'en\'erale de l'hydraulique {\bf 51},
 115 
 (1949); in French.)
 For more complicated types of roughness,
  where roughness elements of many
 different sizes are present (as is commonly the case
 in commercial pipes), the bellies may 
 be absent (see, e.g., the paper by
 Kirshmer cited above) or perhaps present after all
 (see, e.g., C.\ F.\ Colebrook, C.\ M.\ White,
 Proc.\ Roy.\ Soc.\ London  A {\bf 161}, 367 
  (1937)). For the latest experimental data and discussion of this issue
 see  J.\ J.\ Allen,   M.\ A.\ Shockling,   and   A.\ J.\ Smits,  
 {\it Evaluation of a universal transitional resistance diagram for 
 pipes with honed surfaces\/} (to appear in Phys.\ Fluids) and 
  M.\ A.\ Shockling, {\it Turbulent flow in a rough pipe\/},
  MSE Dissertation (Princeton  University, 2005).  }
\bibitem{strickler}{Reprinted in  English in 
A.\ Strickler, {\it Contribution to the question of a velocity 
formula and roughness data for 
streams, channels and close pipelines\/}, translation by 
T.\ Roesgen, W.\ R.\ Brownlie (Caltech, Pasadena, 1981). 
The value $1/3$ of the 
exponent of $r/R$ in Strickler's scaling 
  can be derived by dimensional analysis 
  from the value $2/3$ of the exponent of the 
hydraulic radius in Manning's empirical formula 
for the  average velocity of the flow in a 
  rough open channel. Manning obtained his formula 
independently of Strickler, on the basis of 
different experimental data. 
} 
\bibitem{chow}{See, e.g., V.\ T.\  Chow,
  {\it Open-Channel Hydraulics\/}
  (McGraw-Hill, New York, 1988).}
\bibitem{rich}{L.\ F.\ Richardson,
   Proc.\ Roy.\ Soc.\ London  A {\bf 110}, 709 
 (1926).} 
\bibitem{kolmo}{Reprinted in English in A.\ N.\ Kolmog\'orov,
    Proc.\ R.\ Soc.\ London A {\bf 434}, 9 
    (1991).} 
\bibitem{frisch}{U.\ Frisch, {\it Turbulence\/} 
(Cambridge Univ. Press, Cambridge, 1995).} 
\bibitem{pope}{S.\ B.\ Pope, {\it Turbulent Flows\/} 
(Cambridge Univ. Press, Cambridge, 2000).} 
\bibitem{tl} See, e.g., H.\ Tennekes, J.\ L.\ Lumley,
{\it A First Course in Turbulence\/} (MIT Press, 1972).
\bibitem{taylor}{G.\ I.\ Taylor, 
 Proc.\ Roy.\ Soc.\ London  A {\bf 151}, 421 
  (1935); D.\ Lohse,
   Phys.\ Rev.\ Lett.\ {\bf 73}, 3223 
  (1994).  The existence of an  upper bound on $\varepsilon$ 
 that is independent of the viscosity has been proved
  {\it mathematically\/}; see Doering, Ch.\ R. and P.\ Constantin, 
 Phys.\ Rev.\ Lett.\ {\bf 69}, 1648 (1992). } 
\bibitem{ff}Kolmogorov's four-fifth law reads
 $\overline{u_r^3} = -(4/5)\, \varepsilon \,r$,
 where the left-hand side
 is the third-order structure function.
 Substituting $r=R$ and estimating $|\overline{u_r^3}|=u_R^3$ leads to
 $\varepsilon=(5/4)\, u_R^3/R$, or $\kappa_\varepsilon=5/4$.
 Experimental estimates of $\kappa_\varepsilon$ are O(1);
  see K.\ R.\ Sreenivasan,
Phys. Fluids {\bf 10}, 528 (1998).
\bibitem{ap} R.\ A.\ Antonia and B.\ R.\ Pearson,
 Flow, Turbulence and Combustion {\bf 64}, 95 (2000).
  For comparison, Tennekes and Lumley  give $\kappa_u=0.033$ for
  the atmosphere's turbulent boundary layer \cite{tl}.
\bibitem{gb}{G.\ Gioia, F.\ A.\ Bombardelli, 
   Phys.\ Rev.\ Lett.\ {\bf  88},  014501 (2002). }
\bibitem{aniso}
  {B.\ Knight,  L.\ Sirovich, 
     Phys.\ Rev.\ Lett.\ {\bf 65}, 1356 
    (1990); T.\ S.\ Lundgren, 
     Phys.\ Fluids {\bf 14}, 638 
    (2002); T.\ S.\ Lundgren,
     ibid.\ {\bf 15}, 1024 
    (2003).  Yet the view of anisotropy  
 as a perturbation superposed on the isotropic
 flow  (see, e.g., L.\ Biferale, I.\ Procaccia, 
 {\it Anisotropy in turbulent flows and
    in turbulent transport\/}, arXiv:nlin.CD/0404014 (2004)) might
 break down close to a wall, where  the exponents 
themselves might change (see 
 M.\ Casciola, P.\ Gualtieri, B.\ Jacob, R.\ Piva,
 Phys.\ Rev.\ Lett.\ {\bf 95}, 024503 (2005)). 
 Nevertheless, note (i) that
  the exponent of the second-order structure function 
  appears to change only by 
 little and (ii) that the Kolmog\'orov
exponent does give the exact scalings of Blasius and Strickler, 
whereas any modified exponent would not.}
\bibitem{ng}
The close connection between the turbulent friction and
 the phenomenological spectrum 
  suggests that in turbulence, just as in phase
transitions, large-scale phenomena are 
 direct manifestations of the small-scale statistics.
 The parallel with phase transitions has been
 adduced by N.\ Goldenfeld, who was prompted
 by our results to reanalyze Nikuradse's data
 and conclude that (\ref{efe1}) is consistent with a turbulent
analogue of Widom scaling near critical points.
  See N.\ Goldenfeld,
 {\it Roughness-induced critical phenomena in a turbulent flow\/},
arXiv:cond-mat/0509439 (2005). 
\bibitem{kd}{Our schematic of Fig.~\ref{fig2}
 may seem to resemble the ``d-type roughness''
 of  A.\ E.\ Perry, W.\ H.\ Schofield, and P.\ N.\ Joubert, 
 J.\ Fluid Mech.\ {\bf 37}, 383 (1969). According to 
 these authors, for this type of roughness
  the turbulent friction does not asymptotically 
 approach a constant value at high Re.
 Nevertheless, as pointed out by Jimenez \cite{jimenez},
 the distinction 
  between k-type roughness (including the type of
 roughness in Nikuradse's pipes) and d-type roughness
 appears to have been predicated on limited
  experimental data, and must be regarded
 with caution.
 In any case, the schematic of Fig.~\ref{fig2}
 represents the 
 rough walls of Nikuradse's experiments \cite{belly} and
  does lead to predictions
 that are in accord with those experiments. 
}
\end{thebibliography}

\end{document}